\documentclass[manuscript]{aastex}
\usepackage{rotating,graphicx}

\shorttitle{Effects of Solar Activity on Taylor Scale and
Correlation Scale}

\shortauthors{Zhou, He, \& Wan}

\begin{document}

\title{Effects of Solar Activity on Taylor Scale and Correlation Scale in Solar Wind Magnetic Fluctuations \\}

\author{G. Zhou\altaffilmark{1,2}, H.-Q. He\altaffilmark{1,3,4}, and W. Wan\altaffilmark{1,3,4}}

\altaffiltext{1}{Key Laboratory of Earth and Planetary Physics,
Institute of Geology and Geophysics, Chinese Academy of Sciences,
Beijing 100029, China; hqhe@mail.iggcas.ac.cn}

\altaffiltext{2}{College of Earth and Planetary Sciences, University
of Chinese Academy of Sciences, Beijing 100049, China}

\altaffiltext{3}{Innovation Academy for Earth Science, Chinese
Academy of Sciences, Beijing 100029, China}

\altaffiltext{4}{Beijing National Observatory of Space Environment,
Institute of Geology and Geophysics, Chinese Academy of Sciences,
Beijing 100029, China}

\begin{abstract}
The correlation scale and the Taylor scale are evaluated for
interplanetary magnetic field fluctuations from two-point, single
time correlation function using the Advanced Composition Explorer
(ACE), Wind, and Cluster spacecraft data during the time period from
2001 to 2017, which covers over an entire solar cycle. The
correlation scale and the Taylor scale are respectively compared
with the sunspot number to investigate the effects of solar activity
on the structure of the plasma turbulence. Our studies show that the
Taylor scale increases with the increasing sunspot number, which
indicates that the Taylor scale is positively correlated with the
energy cascade rate, and the correlation coefficient between the
sunspot number and the Taylor scale is 0.92. However, these results
are not consistent with the traditional knowledge in hydrodynamic
dissipation theories. One possible explanation is that in the solar
wind, the fluid approximation fails at the spatial scales near the
dissipation ranges. Therefore, the traditional hydrodynamic
turbulence theory is incomplete for describing the physical nature
of the solar wind turbulence, especially at the spatial scales near
the kinetic dissipation scales.
\end{abstract}


\clearpage

\section{Introduction}
Solar wind turbulence has received considerable attention for
several decades within the community
\citep[e.g.,][]{1965PhFl....8.1385K,1971ApJ...168..509B,1982JGR....87.6011M,1982JGR....8710347M,1995SSRv...73....1T},
and recently there has been an upsurge in interest in this topic.
Much of the interest is driven by the fact that the solar wind can
provide a perfect natural laboratory for the study of plasma
turbulence in space at low-frequency magneto-hydrodynamic (MHD)
scales, which is essential for the studies of solar wind generation,
plasma heating, energetic particle acceleration, cosmic ray
propagation, and space weather
\citep{1968ApJ...152..799J,1968ApJ...152..997J,1968ApJ...153..371C,1979sswp.book..249B,1984PhRvL..53.1449M}.
During the past decades, the solar wind fluctuation properties and
equations of motion have been studied in great detail
\citep{1971ApJ...168..509B,1982JGR....87.6011M,1984JGR....89.9695T,1989JGR....9411739T,1997SoPh..171..363T,1997SoPh..176...87M}.
However, even till now, it is still impossible to make accurate
quantitative predictions. Most of the earlier studies used the
single-spacecraft time-lagged data to infer solar wind spatial
properties based on the well-known ``frozen-in flow" approximation
\citep{1938RSPSA.164..476T}. Provided that the solar wind flow speed
$\textbf{V}_{sw}$ is much greater than the local Alfv{\'e}n speed
$\textbf{V}_{A}$, thus the solar wind fluctuations which pass a
detector are convected in a short time compared to all relevant
characteristic dynamical timescales, so the time lags $\Delta t$ are
equivalent to spatial separations $\Delta t\textbf{V}_{sw}$. This
assumption is relatively reliable under some specific conditions
\citep{1998JGR...10314601P,2000JASTP..62..757R}. However, the time
scale over which the ``frozen-in flow" assumption remains valid is
not fully established
\citep{2005PhRvL..95w1101M,2013JGRA..118.3995W}, and the correct way
to establish the spatial structure is to make use of the
simultaneous two-point single time measurements.

To overcome the shortcomings of the ``frozen-in flow" assumption,
\citet{2005PhRvL..95w1101M} obtained the two-point correlation
function making use of simultaneous two-point measurements of the
magnetic fields. By using this technique, both the correlation
length scale and the Taylor scale can be determined, and the values
of these two scales were given to be 186 $Re$ (Earth radius, 1 $Re$
= 6378 km) and $0.39 \pm 0.11$ $Re$, respectively.
\citet{2007AGUFMSM23A1197W} used the same method and obtained the
Taylor scale values from the data of magnetic field fluctuations in
plasma sheet and solar wind based on the Richardson extrapolation
method. They estimated the Taylor scale in the solar wind as $2400
\pm 100$ km, which agrees with the value given by
\citet{2005PhRvL..95w1101M}. Based on the two-point correlation
measurement technique, \citet{2009JGRA..114.7213W} also obtained the
values of the correlation scale and the Taylor scale by analyzing
the magnetic field data measured from the magnetospheric plasma
sheet and the solar wind. Their results showed that in the solar
wind, the correlation scale along the magnetic field is longer than
that along the perpendicular direction, and the Taylor scale is
independent of the magnetic field directions. These conclusions
indicated that the turbulence is anisotropic. Later,
\citet{2010ApJ...721L..10M} provided a method for estimating the
two-time correlation function and the associated Eulerian
decorrelation timescale in the turbulence. This method can compare
the two-point correlation measurements with the single-point
measurements at corresponding spatial separations, and can be used
for studying the temporal decorrelation of magnetic field
fluctuations in space. Recently, by using the direct two-point
measurements and the ``frozen-in flow" assumption,
\citet{2017ApJ...844L...9C} examined the second-order and
fourth-order structure functions of the magnetic turbulence at very
small scales in the solar wind. Their analysis extended several
familiar statistical results, including the spectral distribution of
energy and the scale-dependent kurtosis, down to the spatial scale
near 6 km, and also clarified the earlier results.

All these applications of the two-point correlation measurements
validate the excellent performance of this technique for studying
the structure of the solar wind turbulence. In the current paper we
determine the values of correlation and Taylor scales of the
interplanetary magnetic field fluctuations based on two-point single
time correlation functions throughout a solar cycle to study the
effects of solar activity on the structure of the plasma turbulence
in the solar wind.

This paper is organized as follows. In Section \ref{Fundamental}, we
give a brief description of some basic concepts of the fluid
turbulence, which can also be used in the field of plasma
turbulence. In Section \ref{Methods}, we provide a detailed
description of the method and the procedure of the two-point
measurements. In Section \ref{Results}, we calculate the Taylor
scale and the correlation scale in different time ranges, and
discuss how the solar activity influences the structure of the
plasma turbulence. Several significant conclusions will be provided
in Section \ref{Conclusions}.

\section{Fundamental Concepts in Fluid Turbulence}
\label{Fundamental}

A turbulent flow should satisfy the Navier-Stokes equation, which is
the momentum evolution of an element of fluid and can be written as
\begin{equation}
\frac{{\partial {\bf{u}}}}{{\partial t}} + {\bf{u}} \cdot \nabla
{\bf{u}} = - \frac{1}{\rho }\nabla p + \nu {\nabla ^2}{\bf{u}}.
\label{eq1}
\end{equation}
Here $\bf{u}$ is the velocity, which is a fluctuating quantity in
time $t$ and space $\bf{x}$, $\nabla$ is the gradient with respect
to $\bf{x}$, $\rho$ is the density, $p$ is the pressure, and $\nu$
is the kinematic viscosity. Note that Equation (\ref{eq1})
corresponds to the incompressible case. Furthermore, Equation
(\ref{eq1}) neglects forces (driving forces, gravity) except
pressure. The strength of the nonlinear convective term ${\bf{u}}
\cdot \nabla {\bf{u}}$ against the dissipative term $\nu {\nabla
^2}{\bf{u}}$ in Equation (\ref{eq1}) can be measured by the
``Reynolds number" which is defined as $R=UL/\nu$, where $U$ and $L$
denote the characteristic flow velocity and the characteristic
length scale (or correlation scale in this study), respectively. The
turbulent flow is characterized by large Reynolds numbers, which
requires that the viscous term should be insignificant in this case.
However, the boundary conditions or initial conditions may make it
impossible to neglect the viscous term everywhere in the flow field.
This can be understood by allowing for the possibility that viscous
effects may be associated with the small length scales. Under the
conditions of large Reynolds numbers, to make the dissipative term
be the same order of the convective term, the viscous term can
survive only by choosing a new length scale $l$. Thus,
\begin{equation}
{U^2}/L = \nu U/{l^2}. \label{eq2}
\end{equation}
From Equation (\ref{eq2}), we can derive that
\begin{equation}
\frac{l}{L} \sim {\left(\frac{\nu }{{UL}}\right)^{1/2}} = {R^{ -
1/2}}. \label{eq3}
\end{equation}
The length scale $l$ is called viscous length which represents the
width of the boundary layer. Thus, at very small length scales, the
viscosity can be effective in smoothing out velocity fluctuations.

Since small-scale motions tend to have small time scales, one can
assume that these motions are statistically independent of the
relatively slow large-scale turbulence. If this assumption makes
sense, the small-scale motion should depend only on the rate at
which it is supplied with energy by the large-scale motion and on
the kinematic viscosity. In accordance with the Kolmogorov's
universal equilibrium theory of the small-scale structure
\citep{1941DoSSR..30..301K,1941DoSSR..32...16K}, the rate of energy
supply should be equal to the rate of dissipation. According to the
dimensional analysis, the amount of the kinetic energy per unit mass
in the turbulent flow is proportional to $U^2$; the rate of transfer
of energy is proportional to $U/L \sim 1/T$, where $T$ denotes the
characteristic transfer time; the kinematic viscosity $\nu$ is
proportional to $U \cdot L$; and the dissipation rate per unit mass
$\varepsilon$, which should be equal to the supply rate, is
proportional to $U^{2}/T \sim U^{3}/L$. With these parameters, we
can obtain length, time, and velocity scales as $\eta \equiv {({\nu
^3}/\varepsilon )^{1/4}}$, $\tau \equiv {(\nu /\varepsilon
)^{1/2}}$, and $v \equiv {(\nu \varepsilon )^{1/4}}$, respectively.
These scales are known as the Kolmogorov microscales of length,
time, and velocity. The Reynolds number formed with $\nu$ and $v$ is
${R_\eta } = \eta v/\nu = 1$. The value ${R_\eta } = 1$ indicates
that the small-scale motion is quite viscous, and that the viscous
dissipation adjusts itself to the energy supply by adjusting length
scales. From the expression
\begin{equation}
\varepsilon \sim U^{3}/L,
\label{eq5}
\end{equation}
we know that the viscous dissipation of energy could be estimated
from the large-scale dynamics that does not involve viscosity. Thus,
the dissipation can be seen as a passive process in the sense that
it proceeds at a rate dominated by the inviscid inertial behavior of
the large eddies. The above expression is one of the cornerstone
assumptions of the classical hydrodynamic turbulence. However, we
should keep in mind that the large eddies only lose a negligible
fraction of their energy compared to direct viscous dissipation
effects. Supposing that the time scale of the energy decay is $L^{2}
/ \nu$, then the energy loss proceeds at a rate of $\nu U^{2}
/L^{2}$, which is very small compared to $U^{3}/L$ if the Reynolds
number $R = UL/\nu  = \frac{{{U^3}/L}}{{\nu {U^2}/{L^2}}}$ is very
large. \citet{1941DoSSR..30..301K,1941DoSSR..32...16K} suggested
that the small-scale structure of turbulence is always approximately
isotropic when the Reynolds number is large enough. This is the
well-known local isotropy theory. In isotropic turbulence, the
dissipation rate can be simply expressed as
\begin{equation}
\varepsilon  \sim \nu \frac{{{U^2}}}{{{\lambda ^2}}},
\label{eq6}
\end{equation}
where $\lambda$ denotes the Taylor scale. Combining Equations
(\ref{eq5}), (\ref{eq6}), and the form of the Kolmogorov microscale,
we can obtain
\begin{equation}
\frac{\lambda }{L} \sim {\left(\frac{{\nu}}{UL}\right)^{1/2}} = {R^{ - 1/2}},\\
\frac{\eta }{L} \sim {\left(\frac{\nu }{{UL}}\right)^{3/4}} = {R^{ - 3/4}},\\
\frac{\lambda }{\eta } \sim {\left(\frac{{UL}}{\nu }\right)^{1/4}} =
{R^{1/4}}. \label{eq7}
\end{equation}
Comparing Equation (\ref{eq3}) with Equation (\ref{eq7}), we can see
that the Taylor scale is related to the viscous dissipation.
Equation (\ref{eq7}) also suggests that for hydrodynamic turbulence
with large Reynolds number, the Taylor scale is larger than the
Kolmogorov microscale.

The length scales $L$, $\eta$ and $\lambda$ mentioned above can
characterise the properties of the flow with high Reynolds numbers.
The correlation scale $L$, also known as ``outer" scale or ``energy
containing" scale, is related to the inertial range of the
turbulence. This parameter represents the size of the largest eddy
in the turbulent flow
\citep{1953tht..book.....B,Tennekes1972A,2000ifd..book.....B}. The
large eddies perform most of the transport of momentum and
contaminants, and the energy input also occurs mainly at large
scales. This correlation scale can be measured by classical methods
based on the Taylor's hypothesis and can be associated with the
first bend-over point in the power spectrum of the turbulent
fluctuations. The Kolmogorov microscale $\eta$ (``inner" scale or
dissipation scale) represents the smallest length scale in the
turbulent flow \citep{1970ApJ...160..745J,Tennekes1972A} and it is
at the end of the inertial range. In the view of traditional
hydrodynamics, the viscosity can be effective in smoothing out
fluctuations and dissipates small-scale energy into heat at very
small length scales (note that in the low-collisionality plasmas,
this situation is less clear). A standard method for identifying the
Kolmogorov microscale is to associate it with the breakpoint at the
high wave number end of the inertial range above which the spectral
index of the power spectral density becomes steeper. The Taylor
scale $\lambda$ is first proposed by \citet{1935RSPSA.151..421T}. It
can be associated with the curvature of the two-point magnetic field
correlation function evaluated at zero separation
\citep{Tennekes1972A,2005PhRvL..95w1101M,2010JGRA..11512250W,2011JGRA..116.8102W,2014JGRA..119.4256C}.
In contrast to the correlation scale and the Kolmogorov microscale,
the Taylor scale does not represent any group of eddy size, but it
can characterise the dissipative effects. Moreover, the Taylor scale
is of the same order of magnitude as the Kolmogorov microscale.
Specifically, the latter is often smaller than the former for
hydrodynamic turbulence with large Reynolds numbers.

An essential characteristic of turbulence is the transfer of energy
across scales. The energy resides mainly at large scales, but it can
be transferred across scales by nonlinear processes, and eventually
it arrives at small scales. The dissipation mechanisms at the small
scales would limit the transfer, dissipate the fluid motions, and
release the heat \citep{1953tht..book.....B,Tennekes1972A}. This is
the so-called ``cascade" process. When the associated Reynolds
number and magnetic Reynolds number are large compared to unity,
this process can be expected in hydrodynamics and in fluid plasma
models such as MHD. In previous studies, the cascade process is
investigated through spectral analysis or structure function
analysis
\citep{1982JGR....87.6011M,1982JGR....8710347M,1994JGR....9911519G,1995SSRv...73....1T,1995ARA&A..33..283G,2004RvMP...76.1015Z}.
Many analysis methods describe the inertial range of scale
properties using the well-known power law of Kolmogorov theory for
fluids \citep{1941DoSSR..30..301K,1941DoSSR..32...16K} and its
variants for plasmas \citep{1965PhFl....8.1385K}. In hydrodynamics,
the inertial range (or the self-similar range) mentioned above is
typically defined extending from the correlation scale (where the
turbulence contains most of the energy) down to the Kolmogorov
microscale. In this work, we use the correlation scale $L$ and the
Taylor scale $\lambda$, instead of the correlation scale $L$ and the
Kolmogorov microscale $\eta$, to describe the properties of the
solar wind turbulence since the Taylor scale can be measured
relatively easily \citep{Tennekes1972A,2005JGRA..110.1205W}.

\section{Methods and Procedures}
\label{Methods}

If the turbulence is homogeneous in space, then the means, variances
and correlation values of the fluctuations are independent of the
choice of origin of the coordinate system
\citep{1953tht..book.....B,Tennekes1972A,1979sswp.book..249B,2000ifd..book.....B}.
For a magnetic field ${\bf{B}}({\bf{x}},t) = {{\bf{B}}_0} +
{\bf{b}}$, the mean is $\left\langle {\bf{B}} \right\rangle  =
{{\bf{B}}_0}$, the fluctuation is ${\bf{b}} = {\bf{B}} -
{{\bf{B}}_0}$, and the variance is ${\sigma ^2} = \left\langle
{{{\left| {\bf{b}} \right|}^2}} \right\rangle $. The two-point
correlation coefficient is
\begin{equation}
R({\bf{r}}) = \frac{1}{{{\sigma ^2}}}\left\langle
{{\bf{b}}({\bf{x}}) \cdot {\bf{b}}({\bf{x}} + {\bf{r}})}
\right\rangle.
\label{eq8}
\end{equation}
Here $\bf{r}$ is the separation of two points $\bf{x}$ and ${\bf{x}}
+ {\bf{r}}$. For homogeneity, $R$ and $\bf{B}_0$ are independent of
position $\bf{x}$, though they may be weakly dependent on position
in reality. The $\left\langle {...} \right\rangle $ denotes an
ensemble average. In homogeneous medium, the ensemble average is
equivalent to a suitably chosen time-averaging procedure. For large
$\left| {\bf{r}} \right|$, the well-behaved turbulence becomes
uncorrelated and $R \to 0$.

The direction-averaged correlation scale is defined as
\citep{1982JGR....87.6011M,2005PhRvL..95w1101M}
\begin{equation}
L = \int_0^\infty  {R(r)} dr.
\label{eq9}
\end{equation}
In addition, the Taylor scale can also be associated with the
curvature of ${R(r)}$ at the origin (see \citet{2005PhRvL..95w1101M}
and \citet{2007AGUFMSM23A1197W} for more details). Strictly speaking
the scale defined in Equation (\ref{eq9}) is the integral scale
which is not necessarily equal to the bendover scale of the
spectrum. However, this will not affect the conclusions in this
work. The discussions of different scales and their relations to
each other can be found in \citet{2020SSRv..216...23S}. A model
correlation function with the correct asymptotic behavior is $R(r)
\sim {e^{-r/L}}$, which has often been used as an approximation tool
for estimating $L$ \citep{1982JGR....87.6011M}. Note that $R(r)=1$
for $r=0$ and $R \to 0$ for $r \to \infty$.

From the equations mentioned above, it is clear that the two-point
correlation coefficient ${R(r)}$ plays an important role in
determining the correlation scale $L$ and the Taylor scale
$\lambda$. In the following, we shall focus on the procedures for
obtaining the ${R(r)}$, $L$ and $\lambda$ from multi-spacecraft
data.

The magnetic field data used in this work was obtained by the
instruments on spacecraft ACE, Wind, and Cluster during the time
period from January, 2001 to December, 2017. Most of the distances
between ACE and Wind spacecraft are in the range of $20-500 Re$, and
the Cluster interspacecraft separations during this period range
from about 100 km to over 10000 km. Since the spacecraft ACE and
Wind orbit the Lagrangian point L1 which is about 1.5 million km
from the Earth and 148.5 million km from the Sun, the information of
the solar wind can be directly obtained by the spacecraft. The
Cluster mission, which consists of four identical spacecraft at
different positions, can provide the three-dimensional measurements
of large-scale and small-scale phenomena in the near-Earth
environment \citep{1997SSRv...79...11E}. Note that the four Cluster
spacecraft are not always in the solar wind. Occasionally, the
Cluster spacecraft are in the Earth's magnetosphere. Therefore, the
data provided by the Cluster mission should be filtered before we
use them.

The first step for investigating the spatial scales in the solar
wind turbulence is to identify the time intervals during which the
spacecraft were immersed in the solar wind. Table \ref{table1} shows
the typical values of several solar wind parameters near 1 AU. As we
can see, the values of the magnetic field magnitude and the plasma
parameters drastically change when the spacecraft travel in and out
of the Earth's magnetosphere. Generally, in the solar wind the
plasma velocity is greater than 200 km/s, and the magnetic field
magnitude is of the order of several $nT$. When the spacecraft fly
into the Earth's magnetosphere, however, the plasma velocity
decreases rapidly, and the magnetic field magnitude increases to
several hundred $nT$. Furthermore, both the plasma number density
and the proton temperature also show typically different values for
different cases, namely, in the solar wind and in the magnetosphere.

\begin{figure}
 \epsscale{1.0}
 \plotone{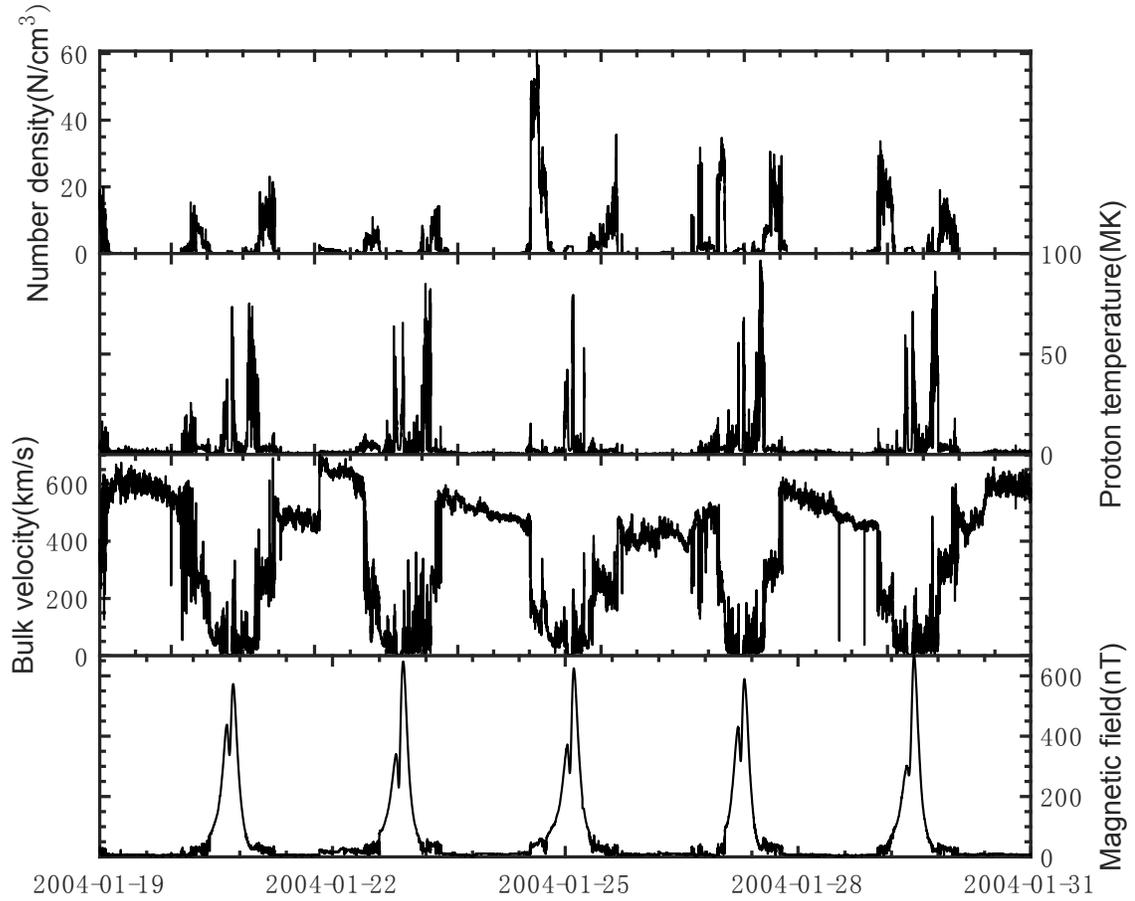}
 \caption{Time series of plasma data measured by FGM and CIS instruments
onboard Cluster 1 during the period January 19-31, 2004. The top,
second, third, and bottom panels denote the particle number density,
proton temperature, solar wind bulk velocity, and magnetic field
magnitude, respectively. \label{fig1}}
\end{figure}

To illustrate the difference in the measurements mentioned above, we
take Figure \ref{fig1} as an example. Figure \ref{fig1} shows the
time series of plasma data measured by Fluxgate Magnetometer (FGM)
and Cluster Ion Spectrometer experiment (CIS) onboard satellite
Cluster 1 during the period January 19-31, 2004. In this time
period, the satellite traveled in and out of the Earth's
magnetosphere about 5 times. We can see the relatively regular
variations of the plasma parameters in Figure \ref{fig1}. When the
satellite crosses the magnetopause, the plasma number density
rapidly increases due to the accumulation of particles there. The
values of the proton temperature and the magnetic field magnitude
also increase, while the plasma bulk velocity sharply decreases.
When the satellite flies out of the magnetopause, however, these
trends are reversed. Based on these behaviors, we can roughly
distinguish the data intervals of the solar wind from those of the
Earth's magnetosphere. In addition, an automated procedure can be
adopted to identify the solar wind intervals. In our investigations,
the solar wind shocks and other discrete solar wind structures are
not removed from the data, since the time period studied is long
enough to neglect the impacts of such solar wind structures.

The measurements from spacecraft ACE and Wind yield two-point
correlation coefficients at larger separations, and those from
spacecraft Cluster provide the correlation coefficients at smaller
separations. For each pair of the spacecraft, we linearly
interpolated the data to 1 min resolution to simultaneously obtain
the field vectors at different spatial positions, since the sampling
rate varied from spacecraft to spacecraft. In order to obtain
meaningful two-point correlation coefficients at larger separations,
longer continuous intervals are required for our analysis.
Therefore, the ACE-Wind data are investigated with a cadence of 1
min, and the individual correlation estimates are calculated by
averaging over contiguous 24 hr periods of data. For Cluster data,
the correlation analysis is carried out with 2 hr sampling.

\begin{figure}
 \epsscale{1.0}
 \plotone{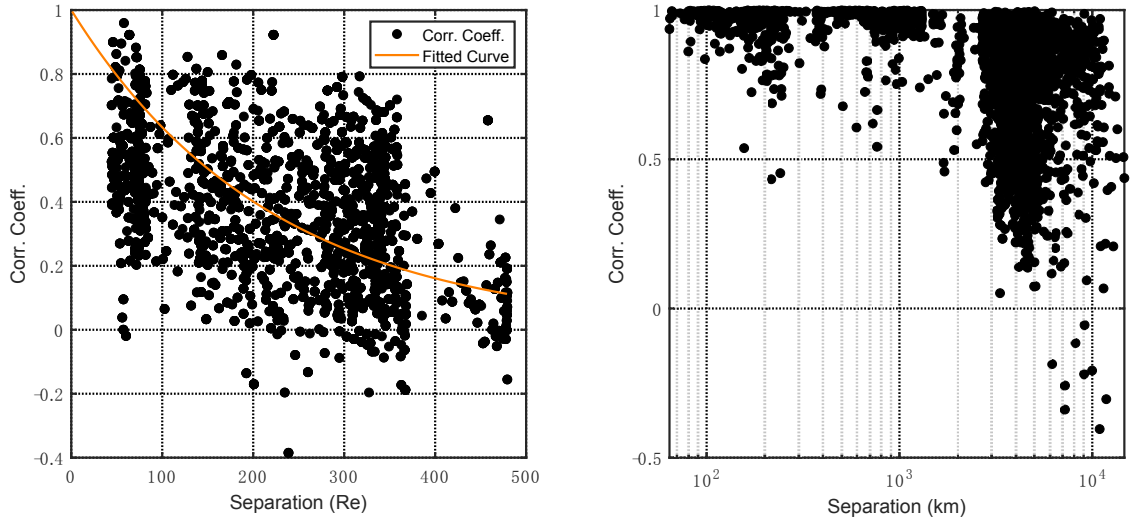}
 \caption{Estimates of solar wind normalized correlation coefficients
 ${R(r)}$ versus spacecraft separations from ACE-Wind data intervals (left) and
Cluster data intervals (right) during the years 2001-2003. The
correlation coefficient for the magnetic field vectors decreases
with the increasing spacecraft separation. Fitting to the ACE-Wind
data (solid curve) in the left panel gives the correlation scale
$L=219Re$. \label{fig2}}
\end{figure}

The data used in this study were measured during the time period
2001-2017 that covers over an entire solar cycle. The spacecraft can
provide us thousands of time intervals for studying the effects of
solar activity on the correlation scale and the Taylor scale. The
entire data set is divided into a series of 3-year time periods. In
each 3-year period, the data intervals are randomly selected. For
each data interval, we calculate the time-averaged two-point
correlation coefficients of the magnetic field vector. The
correlation value is assigned to the time-averaged spacecraft
distance in the corresponding interval. Using the normalized
two-point correlation values calculated from a large number of solar
wind measurements in different divided time periods, we can obtain
the two-dimensional, normalized correlation coefficients as
functions of the spatial separations and the time ranges. For
example, Figure \ref{fig2} shows the estimates of solar wind
correlation coefficients ${R(r)}$ versus spacecraft separations from
ACE-Wind data intervals (left) and Cluster data intervals (right)
during the years 2001-2003. In the left panel of Figure \ref{fig2},
a mean correlation function with the form $R(r) \sim {e^{ - r/L}}$
is obtained by fitting to the data of ACE-Wind correlation
coefficients. Using the definition of the correlation scale $L$
given by \citet{2005PhRvL..95w1101M}, i.e., $R(r) = {e^{ - 1}} =
0.368$, the (direction-averaged) correlation length scale $L$ can be
estimated to be 219 Re during the time period 2001-2003. We can use
the so-called Richardson extrapolation technique (see
\citet{2007AGUFMSM23A1197W} for details) to calculate the Taylor
scale. Using the normalized correlation coefficients from Cluster
data in the right panel of Figure \ref{fig2}, we can obtain that the
Taylor scale is $4311.2 km$ during 2001-2003. Note that the data
intervals are selected with a random procedure. Therefore, the
correlation scale and the Taylor scale may slightly change their
values when we repeat the calculations of them in the same divided
time period. In this work, we use the averaged values of these
repeated calculations for the correlation scale and the Taylor scale
in each time period (2001-2003, 2002-2004, $\ldots$, 2014-2016,
2015-2017).

We can employ the values in different time periods to investigate
the variation trends of the correlation scale and the Taylor scale.
In this work, the sunspot number is chosen to be the indicator of
the solar activity. As we know, the number of sunspots varies with
an 11-year period, which is called the solar cycle
\citep{1979cmft.book.....P,2010LRSP....7....1H}. Generally, more
sunspots indicate that more masses and energies would be released
into interplanetary space through solar burst activities and events.
By means of the data of the sunspot number, the correlation scale,
and the Taylor scale, we can investigate the effects of the solar
activity on the structure of solar wind turbulence.

\section{Results and Discussion}
\label{Results}

\begin{figure}
 \epsscale{1.0}
 \plotone{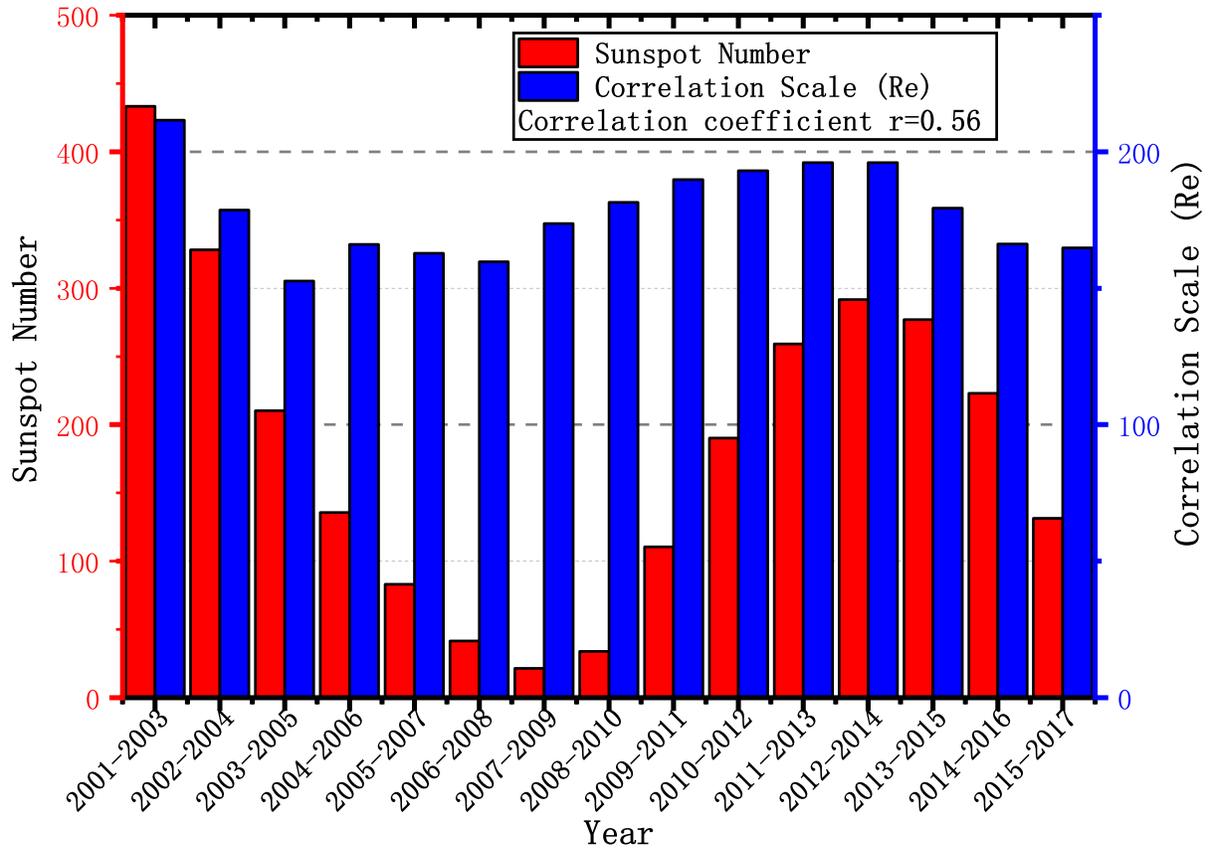}
 \caption{Evolution features of sunspot number (red) and correlation
 scale (blue) during time period 2001-2017. The left and right ordinates
 denote the sunspot number and the correlation scale (Re),
 respectively. The correlation coefficient between the sunspot
 number and the correlation scale is 0.56. \label{fig3}}
\end{figure}

Figure \ref{fig3} displays the evolution of the sunspot number and
the correlation scale during the time period 2001-2017. The left and
right ordinates denote the sunspot number and the correlation scale,
respectively. Obviously, the variation of the sunspots shows a
regular and periodic trend. As we can see, the correlation scale
also shows a weak periodic variation trend. When the sunspot number
is large, the correlation scale becomes large; while when the
sunspot number is small, the correlation scale becomes relatively
small as well. For example, during the time periods 2001-2004 and
2011-2014, the sunspot number and the correlation scale are larger
than those during the time periods 2005-2010 and 2014-2017. As shown
in Figure \ref{fig3}, the maximum and the minimum of the correlation
scale are 211.6 Re and 152.7 Re, respectively. The averaged value of
the correlation scale for all time periods is 178.12 Re, which is
similar to the value 186 Re given in \citet{2005PhRvL..95w1101M}.
The correlation coefficient between the sunspot number and the
correlation scale is 0.56, which suggests a moderate positive
correlation between the solar activity and the correlation scale.
That is to say, the correlation scale of the solar wind turbulence
is modulated by the solar activity to some extent, but not
significantly.

\begin{figure}
 \epsscale{1.0}
 \plotone{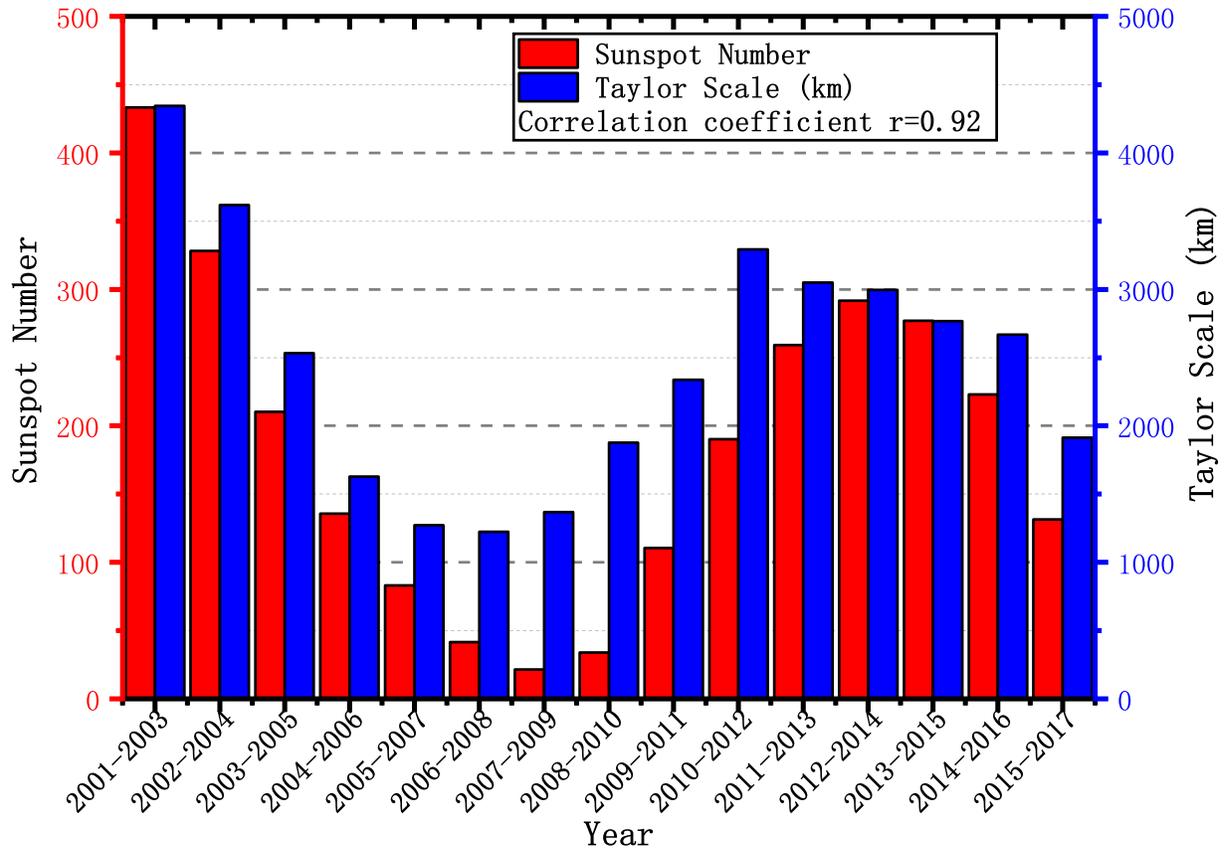}
 \caption{Evolution features of sunspot number (red) and Taylor scale (blue)
 during time period 2001-2017. The left and right ordinates denote the sunspot number and the
Taylor scale (km), respectively. The correlation coefficient between
the sunspot number and the Taylor scale is 0.92. \label{fig4}}
\end{figure}

Figure \ref{fig4} depicts the evolution features of the sunspot
number and the Taylor scale during the time period 2001-2017. The
left and right ordinates denote the sunspot number and the Taylor
scale, respectively. As one can see, relative to the correlation
scale, the Taylor scale is more significantly related to the sunspot
number and the solar activity. The Taylor scale increases or
decreases with the increasing or decreasing sunspot number,
respectively. As shown in Figure \ref{fig4}, the maximum and the
minimum of the Taylor scale are 4345.5 km and 1224.5 km,
respectively. The averaged value of the Taylor scale for all time
periods is 2459.3 km (0.39 Re), which agrees well with the value
$0.39 \pm 0.11$ Re given in \citet{2005PhRvL..95w1101M} and the
value $2400 \pm 100$ km presented in \citet{2007AGUFMSM23A1197W}.
The correlation coefficient between the sunspot number and the
Taylor scale is 0.92, which indicates a strong positive correlation
between the solar activity and the Taylor scale. The high value of
the correlation coefficient means that the Taylor scale is
significantly modulated by the solar activity.

\begin{figure}
 \epsscale{1.0}
 \plotone{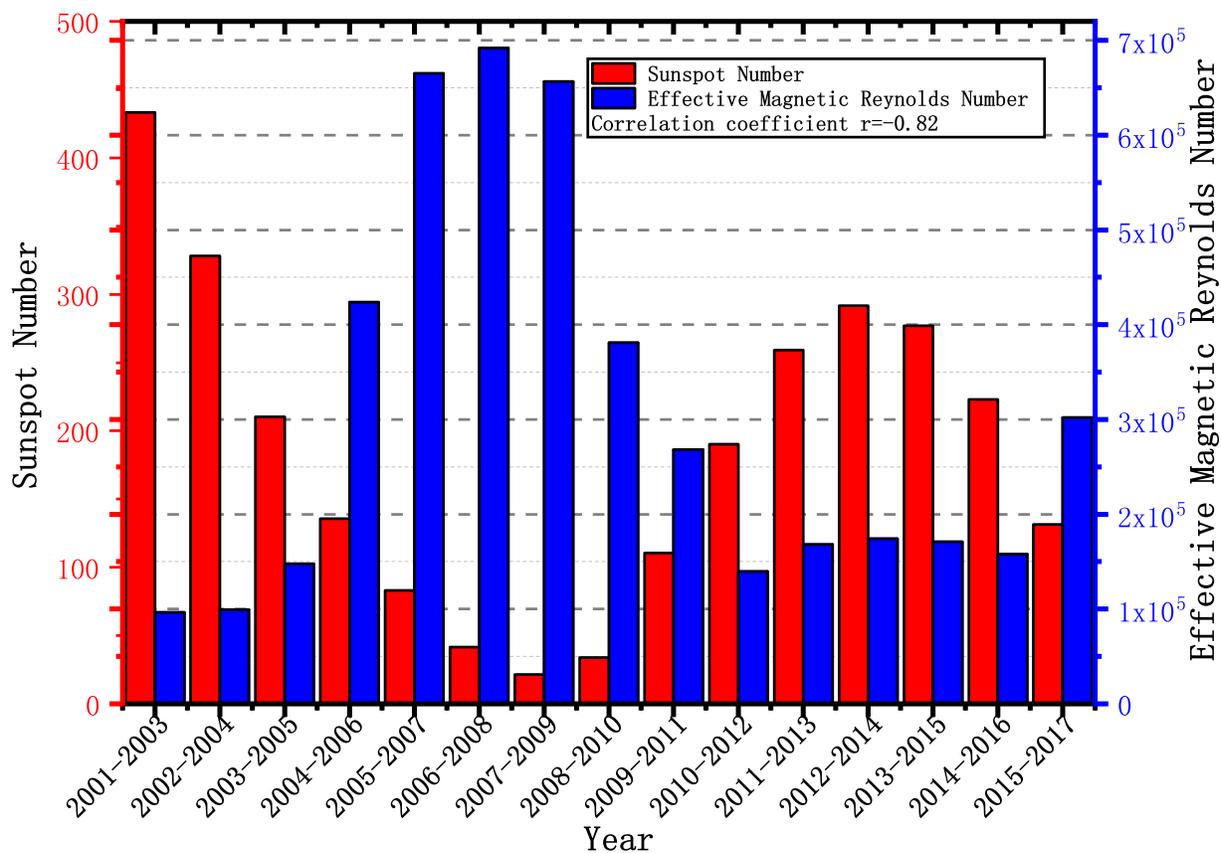}
 \caption{Evolution features of sunspot number (red) and effective magnetic Reynolds
 number (blue) during time period 2001-2017. The left and right ordinates denote the sunspot number and the
effective magnetic Reynolds number, respectively. The correlation
coefficient between the sunspot number and the effective magnetic
Reynolds number is -0.82. \label{fig5}}
\end{figure}

Based on the Equation (\ref{eq7}), we can obtain the form of the
effective magnetic Reynolds number $R_m^{eff}$ as
\begin{equation}
R_m^{eff} = {\left(\frac{L}{\lambda}\right)^2}. \label{eq10}
\end{equation}
Figure \ref{fig5} presents the evolution features of the sunspot
number and the effective magnetic Reynolds number calculated with
Equation (\ref{eq10}) during the time period 2001-2017. The left and
right ordinates denote the sunspot number and the effective magnetic
Reynolds number, respectively. As mentioned above, relative to the
correlation scale, the Taylor scale shows a stronger positive
correlation with the sunspot number. The effective magnetic Reynolds
number shows a negative correlation with the sunspot number. The
correlation coefficient between the sunspot number and the effective
magnetic Reynolds number is -0.82, which indicates that the
turbulence is relatively weak during the time period of strong solar
activity. This result is somewhat counterintuitive.

Generally, the energy output from the Sun varies with the solar
activity. The solar activities include solar flares, coronal mass
ejections (CMEs), extreme ultraviolet emissions, and x-ray emissions
\citep{2010LRSP....7....1H}. Based on the magnetic field data from
the spacecraft ACE and Wind, we have found that both the magnitude
and the standard deviation of the magnetic fields increase during
the rise phase of the solar cycle, and decrease during the declining
phase of the solar cycle. Therefore, the magnetic energy
$\textbf{B}^2$ increases with the increasing solar activity, and
decreases with the decreasing solar activity. Taking into account
the Equation (\ref{eq5}) and replacing the energy with the magnetic
energy, we can know that if the correlation scale $L$ does not
change significantly, the energy dissipation rate $\varepsilon$ will
increase with the increasing magnetic energy $\textbf{B}^2$ during
the rise phase of the solar cycle. Combining this result with the
Equation (\ref{eq6}), we can further derive that in the traditional
theory of hydrodynamic turbulence, the Taylor scale $\lambda$ will
decrease with the increasing $\varepsilon$ during the rise phase of
the solar cycle, and the $\lambda$ will increase with the decreasing
$\varepsilon$ during the declining phase of the solar cycle.
Therefore, according to the traditional theory of hydrodynamic
turbulence, there should be negative correlation between the solar
activity and the Taylor scale. However, our results show that there
is strong positive correlation between them.

This counterintuitive finding is somewhat identical with the results
presented by \citet{2006ApJ...645L..85S} and
\citet{2008ApJ...678L.141M}. In the previous studies, the authors
employed the data sets involving intervals from the magnetic cloud
and noncloud situations in the solar wind to investigate the
spectral properties in the dissipation range. They showed that the
spectral form in the dissipation range is not consistent with the
predictions of the hydrodynamic turbulence and its MHD counterparts.
For instance, the Taylor scale is usually larger than the Kolmogorov
microscale for the hydrodynamic turbulence with large Reynolds
number. However, \citet{2008ApJ...678L.141M} suggested that under
several conditions, the Taylor scale is smaller than the Kolmogorov
microscale even if the magnetic Reynolds number is large enough.
Therefore, the plasma dissipation function is not of the familiar
viscous-resistive Laplacian form. They suggested that the steeper
gradient of the dissipation range spectrum is associated with the
stronger energy cascade rate. For weaker cascade rate, the gradient
of the dissipation range spectrum is gentler. Here, the steep
dissipation range spectrum indicates that the Taylor scale is large.
On the contrary, the gentle dissipation range spectrum means that
the Taylor scale is small. This result is similar to our finding
that there exists positive correlation between the Taylor scale and
the energy dissipation rate. This finding highlights the
non-hydrodynamic properties of the dissipation process in the solar
wind. One possible explanation is that in the solar wind, the
assumption of the fluid approximation fails at the spatial scales
near the dissipation range. Therefore, the traditional hydrodynamic
turbulence theory is incomplete for describing the physical nature
of the solar wind turbulence, especially at the spatial scales near
the kinetic dissipation scales where the particle effects are not
negligible. In the solar wind, the dissipation process of the
turbulence always results from the breakdown of the fluid
approximation and the domination of the kinetic particle effects
such as cyclotron and Landau damping \citep{2006ApJ...645L..85S}.
Therefore, the dissipation process in the solar wind represents the
coupling of the turbulent fluid cascade and the kinetic dissipation.

As the cornerstone assumption of the classical hydrodynamic
turbulence, the viscous dissipation rate of energy can be estimated
by the supply rate of energy at large-scales. This assumption is
described by Equation (\ref{eq5}), and can be modified as:
\begin{equation}
\varepsilon  + \xi  = {U^3}/L. \label{eq11}
\end{equation}
Here $\varepsilon$ is the energy dissipation rate at small scales,
${U^3}/L$ denotes the energy cascade rate, and $\xi$ denotes the
energy loss occurring when the energy transfers from large scales to
small scales especially near the spatial scales at which the solar
wind fluid approximation fails. Equation (\ref{eq11}) indicates that
the sum of the energy dissipation rate and the total energy loss
rate equals to the energy transfer rate. As shown above, the Taylor
scale is strongly positively correlated with the energy cascade
rate. Combining this result with Equation (\ref{eq6}), we can infer
that the Taylor scale $\lambda$ will increase with the increasing
magnetic energy ${{U^2}}$, which indicates that the energy
dissipation rate $\varepsilon$ is relatively stable in the solar
wind turbulence. Therefore, the total energy loss $\xi$ will
increase with the increasing energy cascade rate ${U^3}/L$.
Similarly, the total energy loss $\xi$ will decrease with the
decreasing energy cascade rate ${U^3}/L$. This finding sheds new
light on the relationship between the energy cascade and the
dissipation in the low-collisionality plasma turbulence.

\section{Conclusions}
\label{Conclusions}

In this work, based on the simultaneous measurements (Wind, ACE, and
Cluster) of the interplanetary magnetic fields during the time
period 2001/01-2017/12, we use the two-point, single time
correlation function to determine the fundamental parameters of the
solar wind turbulence, such as the correlation scale and the Taylor
scale. The data set used in this study covers an entire solar cycle.
It is possible to employ this data set accumulated over a long time
period to study the effects of solar activity on the correlation
scale and the Taylor scale. We show that the correlation coefficient
between the sunspot number and the correlation scale is 0.56, and
the correlation coefficient between the sunspot number and the
Taylor scale is 0.92. Obviously, the relationship between the Taylor
scale and the sunspot number is more significant than the
relationship between the correlation scale and the sunspot number.
Therefore, the effective magnetic Reynolds number is primarily
affected by the Taylor scale. The correlation coefficient between
the sunspot number and the effective magnetic Reynolds number is
-0.82, which indicates that the solar wind turbulence is relatively
weak when the solar activity is strong. This result is somewhat
counterintuitive.

In traditional theory of hydrodynamic turbulence, the dissipation
range or the inertial range can be described by a universal
function. The dissipation scale is determined by the energy cascade
rate through the inertial range. Specifically, the stronger cascades
generate the smaller dissipation scales. However, our results
suggest that the form of the dissipation process in solar wind
turbulence is not consistent with the predictions of the
hydrodynamic turbulence and its immediate MHD counterparts. Using
the solar wind data measured at 1 AU, we have shown that the
variation of the Taylor scale depends on the energy cascade rate in
a manner different from the traditional hydrodynamic case. The
Taylor scale increases with the increasing sunspot number, and
decreases with the decreasing sunspot number. This indicates that
the Taylor scale is positively correlated with the energy cascade
rate.

One possible explanation is that in the solar wind, the fluid
approximation fails at the spatial scales near the dissipation
range. Therefore, the traditional theory of hydrodynamic turbulence
is incomplete for describing the physical nature of solar wind
turbulence, especially at the spatial scales near the kinetic
dissipation scale where the particle effects are not negligible. The
dissipation process in the MHD turbulence results from the breakdown
of the fluid approximation and the domination of the kinetic
particle effects such as cyclotron and Landau damping. Therefore,
the dissipation process in the solar wind represents the coupling of
the turbulent fluid cascade and the kinetic dissipation. We suggest
that the energy dissipation rate $\varepsilon$ is relatively stable
in solar wind turbulence. The energy cascade rate ${U^3}/L$ is
positively correlated with the total energy loss $\xi$.

The results presented in this work suggest that solar wind
turbulence is influenced by the solar activity accompanying the
solar cycle. In addition, our investigations highlight the
non-hydrodynamic properties of the dissipation process in the solar
wind, which provides new perspectives on the relationship between
the energy cascade and the dissipation in the low-collisionality
plasma turbulence. The anisotropy of the solar wind turbulence is
another important subject in the field. In the future work, we will
investigate the effects of the solar activities and the solar cycle
on the anisotropy of the solar wind turbulence.


\acknowledgments

This work was supported in part by the B-type Strategic Priority
Program of the Chinese Academy of Sciences under grant XDB41000000,
the National Natural Science Foundation of China under grants
41874207, 41621063, 41474154, and 41204130, and the Chinese Academy
of Sciences under grant KZZD-EW-01-2. H.-Q.H. gratefully
acknowledges the partial support of the Youth Innovation Promotion
Association of the Chinese Academy of Sciences (No. 2017091). We
benefited from the data of ACE, Wind, and Cluster provided by
NASA/Space Physics Data Facility (SPDF)/CDAWeb. The sunspot data
were provided by the World Data Center SILSO, Royal Observatory of
Belgium, Brussels.


\clearpage


\begin{table}
    \caption{Typical values of several solar wind parameters at 1 AU.}
    \centering
    \begin{tabular}{l |c| c| c}
        \hline
        Solar wind parameters  & Minimum values & Maximum values & Mean values\\
        \hline
        Number density  & 0.04 cm$^{-3}$ &  8 cm$^{-3}$ & 5 cm$^{-3}$ \\
        Bulk velocity  & 200 km/s & 900 km/s & 400-500 km/s \\
        Proton temperature  & $5\times10^{3}$ K & $1\times 10^{5}$ K & $2\times 10^{5}$ K \\
        Magnetic field  & 0.25 nT & 40 nT & 6 nT \\
        \hline
    \end{tabular}
    \label{table1}
\end{table}


\clearpage

\clearpage

\clearpage

\clearpage

\clearpage


\end{document}